\documentclass[superscriptaddress,longbibliography]{revtex4-1}
\input{Qcircuit}
\usepackage{amsfonts,amsmath}
\usepackage{filecontents}
\usepackage{siunitx}
\usepackage[letterpaper]{geometry}

\newcommand{\integers}{\mathbb{Z}}
\newcommand{\fastqc}[1]{\Qcircuit @C=1em @R=.7em {#1}\vspace{\baselineskip}}
\newcommand{\parens}[1]{\left(#1\right)}
\newcommand{\braces}[1]{\left[#1\right]}
\newcommand{\floor}[1]{\left\lfloor #1\right\rfloor}

\begin{document}

\title{An $n$-bit general implementation of Shor's quantum period-finding algorithm}
\author{J. T. Davies}
\affiliation{Department of Computer Science, Michigan Technological Unversity, Houghton, MI 49931, USA}
\affiliation{Department of Mathematical
Sciences, Michigan Technological Unversity, Houghton, MI 49931, USA}
\author{Christopher J. Rickerd}
\affiliation{Department of Electrical and Computer
Engineering, Michigan Technological Unversity, Houghton, MI 49931, USA}
\author{Mike A. Grimes}
\affiliation{Department of Electrical and Computer
Engineering, Michigan Technological Unversity, Houghton, MI 49931, USA}
\author{Durdu \"{O}. G\"{u}ney}
\affiliation{Department of Electrical and Computer
Engineering, Michigan Technological Unversity, Houghton, MI 49931, USA}

\begin{abstract}
   The goal of this paper is to outline a general-purpose scalable
   implementation of Shor's period-finding algorithm using fundamental quantum
   gates, and to act as a blueprint for linear optical implementations of
   Shor's algorithm for both general and specific values of $N$. This offers
   a broader view of a problem often overlooked in favour of compiled versions
   of the algorithm.
\end{abstract}

\maketitle

\section{Introduction}
The superiority of quantum computers over classical computers for problems 
with solutions based on quantum Fourier transform, as well as database search 
and quantum simulations continue to attract attention. There are 
many approaches proposed for building scalable quantum computers or achieving long-distance and high data-rate quantum 
communications. Major model physical systems include nuclear magnetic 
resonance, ion trap, neutral atom, cavity QED, solid state, superconducting, 
and optical approaches. All of these have their own advantages, but 
unfortunately, also their own drawbacks. None of them on its own is shown to 
be self-sufficient to build a large-scale practical quantum computer or quantum  networks. 

Quantum photonic integrated circuits, based on semiconductor technology and 
compatibility with existing microelectronics infrastructure, have been recently 
envisioned as the route to utilizable quantum information technologies 
enabling robust and compact quantum circuit boards and processors of the next 
generation computers and networking devices \cite{c36,c37,c31,c32,c39,c35,c38}.
However, thousands of stable, interconnected interferometers with low-loss and high 
performance components are unavoidable for the implementation of practical 
large-scale quantum algorithms, since quantum error correcting codes are not 
very helpful at that large scale \cite{c37,c35}. Therefore, bulky optical 
experiments with only several photons have already started to move toward 
stable, miniaturized integrated quantum circuits with many logic gates like 
at the heart of classical computers \cite{c36,c37,c35,c38}.

Silica-on-silicon waveguides \cite{c5,c1,c2,c3,c10,c19,c20}, silicon-on
insulator waveguides \cite{c16,c22,c18,c34}, GaN-on-sapphire waveguides
\cite{c14}, direct laser writing \cite{c2,c4,c7,c8,c9,c20,c21,c24}, and
standard lithography techniques \cite{c5,c1,c14,c10,c19,c16,c18,c22,c34} were
chosen as underlying technologies for the implementations of quantum photonic
integrated circuits. Non-classical interference of photon pairs in integrated
optical circuits using a phase controlled Mach-Zehnder interferometer
\cite{c2}, directional coupler \cite{c5, c16}, and integrated two
four-wave mixing waveguide sources in a phase controlled interferometer
\cite{c18} was shown. In 2008, Politi, et al. \cite{c5} demonstrated integrated
optical controlled-NOT gate and path-entanglement. The group later
implemented Shor's quantum algorithm on a photonic chip to factorize 15
\cite{c1}, heralded path-entangled NOON states up to four photons for
high precision quantum metrology \cite{c3}, reconfigurable photonic circuits
for the generation and manipulation of entangled states \cite{c6,c10,c34}, and
manipulation of externally generated single photons \cite{c19}. Sansoni, et al.
\cite{c7} reported the realization of a directional coupler functioning as a
beam splitter for polarization qubits. This work was then extended to
demonstrate the first integrated photonic controlled-NOT gate for polarization
qubits \cite{c8}. Corrielli, et al. showed the capability of performing
arbitrary transformations on polarization qubits in an integrated waveguide
circuit \cite{c24}. Implementations of quantum walk using integrated photonic
waveguides were proposed for potential applications in quantum simulations
\cite{c4, c9}. Single photon detectors with 20\% efficiency \cite{c11} and 
photon-number-resolving detectors with 24\% efficiency \cite{c26} using 
superconducting nanowires on GaAs ridge waveguides and single-mode waveguide 
photon-number-resolving detector with 40\% efficiency \cite{c20} were shown at
telecom wavelengths for photonic quantum circuits. Najafi, et al. \cite{c28}
demonstrated on-chip detection of externally generated entangled photons by 
integrating multiple superconducting nanowire single photon detectors. Mower 
and Englund proposed a theoretical protocol for on-demand generation of 
single and entangled photons on a silicon photonic integrated chip using a 
time-multiplexed spontaneous parametric down conversion element \cite{c12}. In 
2013, Spring, et al. showed the first single photon source on a silica 
photonic chip based on spontaneous four wave mixing with a heralding 
efficiency of 40\% \cite{c21}. Spatial multiplexing of heralded single photon 
sources on monolithic silicon chip using spontaneous four wave mixing in 
photonic crystal waveguides was proposed to increase the heralded photon rate 
\cite{c22}. Matsuda, et al. \cite{c29} reported generation and demultiplexing
of photon pairs on a silicon-silica monolithic waveguide platform. Based on
type II-phase matched spontaneous parametric down-conversion processes,
experimental polarization entangled post-selection free photon source
\cite{c25} and theoretical hyperentangled photon pair generation \cite{c15}
were shown. Takesue, et al. \cite{c27} reported an on-chip single-photon
buffering for 150ps based on coupled resonator optical waveguides consisting
of high-Q photonic crystal cavities. Mouradian, et al. \cite{c33} demonstrated
quantum memories with \SI{120}{\micro\second} spin coherence times based on
nitrogen-vacancy centers in a photonic circuit. Metcalf, et al. \cite{c30}
have very recently reported the first experimental demonstration of quantum
teleportation on a reconfigurable integrated photonic chip which performs
entanglement preparation, Bell-state analysis, and quantum state tomography,
similar to the quantum photonic crystal integrated circuits that we envisioned
in 2007 \cite{c31, c32}. Li, et al. \cite{c23} have proposed an experimental
calibration method for quantum photonic integrated circuits that are becoming
increasingly complex. Tezak, et al. \cite{c13} proposed a quantum hardware
description language to facilitate the analysis, design, and simulation of
complex photonic circuits.  

Despite all the promising developments above in theory and experiments, 
desired level of progress in realization of quantum circuits and algorithms 
has not been yet achieved. It is essential to explore large-scale integration 
of high quality single and entangled photon sources on demand, long-lived 
quantum memories, high-efficiency single photon detectors, reconfigurable 
implementations of quantum logic gates and protocols.

Here, we construct a general implementation of Shor's 
algorithm for any appropriate n-bit number using quantum logic gates, which 
can then be easily translated into quantum photonic integrated circuits based 
on directional couplers, interferometers, single-photon sources, detectors, 
and other integrated optical devices. This general representation of Shor's 
algorithm, as a model, in the physical layer can provide a simple recipe for 
both experimentalists and theorists toward large-scale practical 
implementations and help foresee the technological and fundamental limits of 
photonic integrated circuit approach to determine and focus on the most 
important experimental and theoretical requirements and aspects that are 
often overlooked. 

Shor's quantum factorization algorithm provides a means to easily factor 
integers that are the product of two primes. Such numbers, and the difficulty 
of factorizing them, provide the basis of the common RSA encryption 
algorithm; a polynomial-time factorization method would greatly impact the
effectiveness of this algorithm, with significant influence on the development
of software, particularly when dealing with Internet security.

In section II, we introduce the basic design of Shor's period-finding quantum
circuit and its component modules. In section III, we begin developing basic
arithmetic operations on qubit registers; these are used in section IV to
build low-level modular operations. Section V uses this block to build the
operations necessary to perform modular arithmetic, while section VI details the
use of these block to produce the modular exponentiation module needed by
Shor's period-finding circuit. We conclude in section VII.

Unless otherwise stated, we assume that upper-case variables ($A$, $B$) are $n$-
bit numbers stored in $n+1$-bit registers (the top bit is always a 0, to have 
room for addition/multiplication by 2), and lower-case variables are single 
bits. A lower-case variable with a subscript indicates a specific bit of a 
number, indexed from zero; for example, $a_{2}$ is the third bit of $A$.

\section{Shor's Factorization Algorithm and its Quantum Circuit Representation}
Shor's factorization algorithm contains both classical and quantum processes.
The only section of the algorithm considered to be non-classical consists of a
quantum circuit designed to find the period of a modular exponentiation
function. This circuit can be further broken down into three modules. The
complete quantum circuit is depicted below, using three black boxes:

$$\fastqc{
   \lstick{0} &            \qw &          \qw & \multigate{4}{A^{Y}\%N} & 
      \rstick{0} \qw\\
   \lstick{1} &            \qw &          \qw &        \ghost{A^{Y}\%N} &
      \push{A^{Y}\% N} \qw & \meter\\
   \lstick{N} &            \qw &          \qw &        \ghost{A^{Y}\%N} &
      \rstick{N} \qw\\
   \lstick{A} &            \qw &          \qw &        \ghost{A^{Y}\%N} &
      \rstick{A^{2^{n}}\% N} \qw\\
   \lstick{0} & \gate{H_{n+1}} & \push{Y} \qw &        \ghost{A^{Y}\%N} & 
      \push{Y} \qw & \gate{{QFT}^{-1}} & \qw & \meter\\
	  & (1.) & & (2.) & & (3.)
}$$

There are two numeric inputs to the circuit: $N$, the number to be factored,
and $A$, some number coprime to $N$ chosen in the classical portion of the
algorithm. All of the qubit registers are $n+1$-bits, where $n$ is the number
of bits needed to represent $N$; that is, $n=\floor{\log_{2}N}+1$. 

The first module is a Hadamard transform. Each qubit in the first register is 
subjected to a Hadamard gate. Since this register is initialized to zero, the 
first register is then in a superposition over all states. The number of gates
required for this module scales linearly as the number of qubits increases.

The second module takes three values, $N$, $A$, and $Y$, and returns $A^{Y}\%N$
where $\%$ denotes the modulo operator. It also requires a number of ancillary
qubits beyond those listed, on the order of $n^{2}$. Note that the
$A^{2^{n}}\%N$ ancillary result comes only from known values and those from the
classical portion of the process, and the 0 register and $N$ registers do not
change, so they can be reclaimed for other uses after the algorithm is
finished.

The third module is an inverse quantum fourier transform, or inverse QFT. A
general form for this component is well-known, and the number of gates scales
quadratically with $n$.

Current efforts at realizing Shor's algorithm using photonic gates has 
focused on proof-of-concept methods for specific values, such as 15
\cite{c1} or 21\cite{factor21}. While the first and third modules are well-known
and easily constructed, the modular exponentiation algorithm is more difficult
to construct for a general case. Many experimental attempts to realise Shor's
Algorithm rely on a version of this process optimised for specific constants,
as in \cite{Markov}. These ``compiled'' circuits, while useful for experimental
purposes, are restricted to a small subset of $A$ and $N$ values, and often
obfuscate the nature of the modular exponentiation component. We will focus on
creating the modular exponentiation component for a general case from elementary
gates in a way that is accessible to experimentalists.

\section{Basic Arithmetic Blocks}
\subsection{CDKM Adder $\braces{+}$}
The Cuccaro-Draper-Kutin-Moulton (CDKM) adder is a two's complement
reversible ripple-carry adder that uses no ancillary qubits (counting the 
carry bit as part of the output) \cite{Cuccaro}. It acts as the basis of many 
of the blocks we describe later, and itself consists of two blocks: majority 
and unmajority-and-sum. Several optimisations for the design have emerged
\cite{Thomsen}, but we will be focused on the simplest implementation.

\subsubsection{Majority $\braces{MAJ}$}
The majority block computes $c_{i}\oplus a_{i}$, $a_{i}\oplus b_{i}$, and
$c_{i+1}$ from $c_{i}$, $a_{i}$, and $b_{i}$, where $c_{i}$ is the $i$th carry
value, $a_{i}$ is the $i$th bit of $A$, and $b_{i}$ is the $i$th bit of $B$. 
The majority block is implemented using elemetary gates:

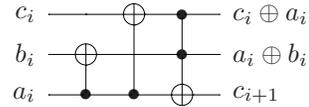
$$\fastqc{
   \lstick{c_{i}} &       \qw &     \targ & \ctrl{1} &
      \rstick{c_{i}\oplus a_{i}} \qw \\
   \lstick{b_{i}} &     \targ &       \qw & \ctrl{1} &
      \rstick{a_{i}\oplus b_{i}} \qw \\
   \lstick{a_{i}} & \ctrl{-1} & \ctrl{-2} &    \targ &
      \rstick{c_{i+1}} \qw
}$$

\subsubsection{Unmajority and Sum $\braces{UMS}$}
The unmajority-and-sum block computes $c_{i}$, $s_{i}$, and $a_{i}$ from
$c_{i}\oplus a_{i}$, $a_{i}\oplus b_{i}$, and $c_{i+1}$; that is, from the 
outputs of the corresponding majority block.

$$\fastqc{
   \lstick{c_{i}\oplus a_{i}} & \ctrl{1} &     \targ & \ctrl{1} &
      \rstick{c_{i}} \qw \\
   \lstick{a_{i}\oplus b_{i}} & \ctrl{1} &       \qw &    \targ &
      \rstick{s_{i}} \qw \\
             \lstick{c_{i+1}} &    \targ & \ctrl{-2} &      \qw &
      \rstick{a_{i}} \qw
}
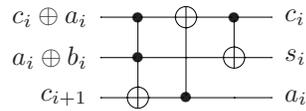
$$

\cite{Cuccaro} offers an alternative version of the UMS block; for ease of 
implementing the controlled adder, we have selected the simpler implementation.

\subsubsection{The Complete Adder}
A pair of these blocks is needed for each bit to be added; for example, a 
CDKM adder adding three bits of input would be
$$\fastqc{
   \lstick{0} & \multigate{2}{MAJ} &                \qw &                
       \qw &      \qw &                \qw &                \qw &
       \multigate{2}{UMS} &     \rstick{0} \qw\\
   \lstick{b_{0}} &        \ghost{MAJ} &                \qw &                
      \qw &      \qw &                \qw &                \qw &
      \ghost{UMS} & \rstick{s_{0}} \qw\\
   \lstick{a_{0}} &        \ghost{MAJ} & \multigate{2}{MAJ} &                
      \qw &      \qw &                \qw & \multigate{2}{UMS} &
      \ghost{UMS} & \rstick{a_{0}} \qw\\
   \lstick{b_{1}} &                \qw &        \ghost{MAJ} &                
   \qw &      \qw &                \qw &        \ghost{UMS} &                
   \qw & \rstick{s_{1}} \qw\\
   \lstick{a_{1}} &                \qw &        \ghost{MAJ} &
   \multigate{2}{MAJ} &      \qw & \multigate{2}{UMS} & \ghost{UMS} &\qw & 
   \rstick{a_{1}} \qw\\
   \lstick{b_{2}} &                \qw &                \qw & 
      \ghost{MAJ} &      \qw &        \ghost{UMS} &                \qw &
         \qw & \rstick{s_{2}} \qw\\
   \lstick{a_{2}} &                \qw &                \qw & 
      \ghost{MAJ} & \ctrl{1} &        \ghost{UMS} &                \qw &
      \qw & \rstick{a_{2}} \qw\\
   \lstick{0} &                \qw &                \qw &                
      \qw &    \targ &                \qw &                \qw &
      \qw &     \rstick{c} \qw
}$$
where $c$ denotes the carry bit of the sum. The carry bit is omitted in many 
of our uses of the CDKM adder; as one of the addends is preserved, no 
information is destroyed by not passing on a carry bit.

\subsection{Controlled CDKM Adder $\braces{+_{C}}$}
A cursory examination of the MAJ and UMS blocks reveals the means to make a 
controlled CDKM adder: placing the blocks side-by-side gives

$$\fastqc{
   \lstick{c_{i}} &       \qw &     \targ & \ctrl{1} & \ctrl{1} &     \targ & 
      \ctrl{1} & \rstick{c_{i}} \qw \\
   \lstick{b_{i}} &     \targ &       \qw & \ctrl{1} & \ctrl{1} &       \qw &
      \targ & \rstick{s_{i}} \qw \\
   \lstick{a_{i}} & \ctrl{-1} & \ctrl{-2} &    \targ &    \targ & \ctrl{-2} &
      \qw & \rstick{a_{i}} \qw
}$$

Consider the following modification:
$$\fastqc{
   \lstick{c_{i}} &       \qw &     \targ & \ctrl{1} & \ctrl{1} &     \targ &
      \ctrl{1} & \rstick{c_{i}} \qw \\
   \lstick{b_{i}} &     \targ &       \qw & \ctrl{1} & \ctrl{1} &       \qw &
      \targ & \rstick{s_{i}} \qw \\
   \lstick{a_{i}} & \ctrl{-1} & \ctrl{-2} &    \targ &    \targ &
      \ctrl{-2}\qw & \rstick{a_{i}} \qw \\
   \lstick{x} & \ctrl{-1} &       \qw &      \qw &      \qw &       \qw & 
      \ctrl{-2} &     \rstick{x} \qw
}$$

If bit $x$ is high, the gates behave as usual; otherwise, the Toffoli gates 
do nothing, leaving the central gates to cancel each other out. This disables 
a single pair of MAJ/UMJ blocks; hooking the control bit $x$ to every other 
pair will similarly disable the entire adder \cite{Markov}. Calling the 
modified blocks CMJ and CUS, the controlled CDKM adder is then
$$\fastqc{
       \lstick{0} & \multigate{2}{CMJ} &                \qw &             \qw &
       \qw &                \qw &                \qw & \multigate{2}{CUS} &
       \rstick{0} \qw\\
   \lstick{b_{0}} &        \ghost{CMJ} &                \qw &             \qw &
   \qw &                \qw &                \qw &        \ghost{CUS} &
   \rstick{s_{0}} \qw\\
   \lstick{a_{0}} &        \ghost{CMJ} & \multigate{2}{CMJ} &             \qw &
   \qw &                \qw & \multigate{2}{CUS} &        \ghost{CUS} &
   \rstick{a_{0}} \qw\\
   \lstick{b_{1}} &                \qw &        \ghost{CMJ} &             \qw &
   \qw &                \qw &        \ghost{CUS} &                \qw &
   \rstick{s_{1}} \qw\\
   \lstick{a_{1}} &                \qw &     \ghost{CMJ} & \multigate{2}{CMJ} &
   \qw & \multigate{2}{CUS} &        \ghost{CUS} &                \qw &
   \rstick{a_{1}} \qw\\
   \lstick{b_{2}} &                \qw &                \qw &     \ghost{CMJ} &
   \qw &        \ghost{CUS} &                \qw &                \qw &
   \rstick{s_{2}} \qw\\
   \lstick{a_{2}} &                \qw &                \qw &     \ghost{CMJ} &
   \ctrl{1} &        \ghost{CUS} &                \qw &                \qw &
   \rstick{a_{2}} \qw\\
       \lstick{0} &                \qw &                \qw &             \qw &
       \targ &                \qw &                \qw &                \qw &
       \rstick{c} \qw\\
       \lstick{x} &          \ctrl{-6} &          \ctrl{-4} &       \ctrl{-2} &
       \qw &          \ctrl{-2} &          \ctrl{-4} &          \ctrl{-6} &
       \rstick{x} \qw
}$$

The output of this gate is
$$\fastqc{
   \lstick{B} & \multigate{1}{+_{C}} & \rstick{B+Ax} \qw\\
   \lstick{A} &        \ghost{+_{C}} &    \rstick{A} \qw\\
   \lstick{x} &            \ctrl{-1} &    \rstick{x} \qw
}$$

\subsection{Alternative Addition Blocks}
The CDKM adder presented above is the most intuitive implementation of a
quantum adder, but it is not necessarily the best when used in an actual
implementation. Each MAJ and UMS block uses three quantum gates, which means
$6n$ gates are needed for an $n$-bit adder, and each result relies on the
previous to be computed, so result takes a long time to compute.

Thomsen and Axelsen\cite{Thomsen} offer an optimisation of the ripple-carry
adder that divides a task of $ck$ bits into $c$ independent $k$-bit adders,
which are then combined to give the result in a total of $O\parens{c+k}$ time.
The Thomsen-Axelsen adder, while offering a substantial speedup over a
$ck$-bit CDKM adder, uses substantially more gates: the hardware cost of the
Thomsen-Axelsen adder is $O\parens{n\sqrt{n}}$, while the CKDM adder is linear.
The improvement in runtime is greater asymptotically than the increase in
hardware cost, so substituting the Thomsen-Axelsen adder should for large $n$
offer an overall improvement, but the higher hardware cost could prove
prohibitive for implementations where the expense of added hardware outweighs
the need for fast calculations.

Draper, Kutin, Rains, and Svore\cite{dkrs} propose a carry-lookahead adder
which runs in $O\parens{\log n}$ time, but requires $O\parens{n}$ ancillary
qubits. The additional qubit cost substantially increases the overhead qubits
required for modular exponentiation.

Zalka uses parallelised addition in \cite{zalka2008} to achieve a reduction of
depth of modular exponentiation from $n^{3}$ to $n^{2}$, while remaining in
$O\parens{n^{3}}$ gates and $O\parens{n}$ qubits. We do not consider
parallelisation of Shor's algorithm extensively in this paper, as it
complicates circuit design, but in a physical implementation the change to a
parallel adder can be advantageous.

\subsection{CDKM Subtractor $\braces{-}$}
As it is simply a two's-complement adder, it is simple to adapt a CDKM adder 
into a CDKM subtractor. The NOT gate is an inherently reversible operation; 
as we know our inputs will be $n$-bit positive integers, the last bit will 
always be 0, and thus there is no need for an ancillary carry bit to add 1. A 
reversible, $n+1$-bit ``add one'' block for $n$-bit integers is simply a CDKM 
adder that substitutes 1 for $A$:

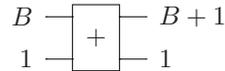
$$\fastqc{
   \lstick{B} & \multigate{1}{+} & \rstick{B+1} \qw\\
   \lstick{1} &        \ghost{+} &   \rstick{1} \qw
}$$

This requires an entire additional $n+1$-bit register, which is inefficient; 
however, we can note that the values of this register are constant across all 
implementations, and as such can determine the inputs of the $MAJ$ gates. The 
lowest-order bit has 1 added and a fixed initial carry of 0, giving
$$\fastqc{
       \lstick{0} &       \qw &     \targ & \ctrl{1} &    \rstick{1 \oplus 0}
         \qw \\
   \lstick{b_{0}} &     \targ &       \qw & \ctrl{1} & \rstick{b_{0}\oplus 1}\\
       \lstick{1} & \ctrl{-1} & \ctrl{-2} &    \targ &         \rstick{c_{1}}
}$$
which is equivalent to
$$\fastqc{
       \lstick{0} &   \qw & \targ &      \qw &      \rstick{1} \qw \\
   \lstick{b_{0}} & \targ &   \qw & \ctrl{1} & \rstick{\neg b_{0}} \qw\\
       \lstick{1} &   \qw &   \qw &    \targ & \rstick{b_{0}} \qw \\
}$$

The subtract one block is similar, but with the other register initialised to 
-1 rather than 1. The number of gates cannot be reduced as easily as with the
+1 gate, since all it does is replace CNOT with NOT in the MAJ block.

Combining these with an $n+1$-bit CDKM adder gives an $n+1$-bit CDKM 
subtractor:
$$\fastqc{
   \lstick{A} &   \qw &       \qw & \multigate{1}{+} &       \qw &   \qw &
      \rstick{A-B} \qw\\
   \lstick{B} & \targ & \gate{+1} &        \ghost{+} & \gate{-1} & \targ &
      \rstick{B} \qw
}$$

\subsubsection{Controlled CDKM Subtractor $\braces{-_{C}}$}
Creating a controlled subtractor can be accomplished by modifying the regular 
subtractor: replace each of the components with a controlled equivalent. 
Controlled NOT is an elementary gate, while the +1 block can be replaced by 
an ordinary CDKM adder\cite{Cuccaro}:

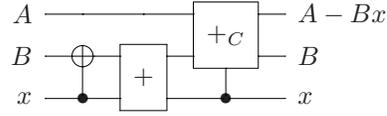
$$\fastqc{
   \lstick{A} &       \qw &              \qw & \multigate{1}{+_{C}} &
      \rstick{A-Bx} \qw \\
   \lstick{B} &     \targ & \multigate{1}{+} &        \ghost{+_{C}} &
      \rstick{B} \qw \\
   \lstick{x} & \ctrl{-1} &        \ghost{+} &            \ctrl{-1} &
      \rstick{x} \qw
}$$

Note that while the +1 block can be manipulated due to not assuming the value
of the first bit, the -1 block must be changed into a controlled version
directly.

\subsection{CDKM Greater-Equal Comparator $\braces{\ge}$}
To determine if $a\ge N$, we introduce one more modification to the CDKM 
adder, the CDKM comparator. The comparator is similar to a subtractor, but 
replaces unmajority-and-sum blocks with simple unmajority ($\braces{UMJ}$) 
blocks, which are the exact opposite of majority blocks:
$$\fastqc{
   \lstick{c_{i}\oplus a_{i}} & \ctrl{1} &     \targ &       \qw &
      \rstick{c_{i}} \qw \\
   \lstick{a_{i}\oplus b_{i}} & \ctrl{1} &       \qw &     \targ &
      \rstick{b_{i}} \qw \\
             \lstick{c_{i+1}} &    \targ & \ctrl{-2} & \ctrl{-1} &
         \rstick{a_{i}} \qw
}$$

Replacing a CDKM adder's UMS blocks with UMJ blocks changes the output so 
that it returns the original inputs and the carry bit only; the sum is not 
returned. We shall refer to this as the comparator base block, $\braces{CMB}$.
$$\fastqc{
   \lstick{A} & \multigate{2}{CMB} & \rstick{A} \qw \\
   \lstick{B} &        \ghost{CMB} & \rstick{B} \qw \\
   \lstick{0} &        \ghost{CMB} & \rstick{c} \qw
}$$

In a subtractor, this will simply return $A$ and $-B$, and the carry qubit 
will be high if there was a carry from the
subtraction:
$$\fastqc{
   \lstick{A} &   \qw &       \qw & \multigate{2}{CMB} &       \qw &   \qw &
      \rstick{A} \qw \\
   \lstick{B} & \targ & \gate{+1} &        \ghost{CMB} & \gate{-1} & \targ &
      \rstick{B} \qw \\
   \lstick{0} &   \qw &       \qw &        \ghost{CMB} &       \qw &   \qw &
      \rstick{c} \qw
}$$

This carry will only be high if $B$ is strictly greater than $A$; it will be 
0 if $B=A$. Thus, we want to know if $B+1$ is strictly greater than $A$; if $B
<A$, then $B+1\le A$ and the carry qubit will be low, while if $B\ge A$, $B+1>
A$ and the bit is high. Thus, by adding one last +1 block, we obtain the 
greater-equal form of the CDKM comparator:

$$\fastqc{
   \lstick{A} &       \qw &   \qw &       \qw & \multigate{2}{CMB} &       \qw&
      \qw & \rstick{A} \qw \\
   \lstick{B} & \gate{+1} & \targ & \gate{+1} &        \ghost{CMB} & \gate{-1}&
      \targ & \rstick{B} \qw \\
   \lstick{0} &       \qw &   \qw &       \qw &        \ghost{CMB} &       \qw&
      \qw & \rstick{c} \qw
}$$

\subsubsection{Other Comparator Implementations}
It is possible to implement a faster, but more complex, comparator by modifying
a Thomsen-Axelsen adder in a similar way to the modification of the CDKM adder,
with the corresponding performance increases; simply replace the block
unmajority and sum with block unmajority, and extract the maximum carry with a
CNOT gate and an empty qubit.

\subsection{Doubling Block $\braces{\times 2}$}
Our implementation of multiplication necessitates the ability to multiply the
input $n$-bit numbers by two. Recall that all of the registers are $n+1$-bit;
as we know that $N$ is an $n$-bit number, $A$ cannot be more than $n$ bits.
Thus, a simple series of swaps can be used to multiply $A$ by two. For
instance, in the case where $n=5$,
$$\fastqc{
       \lstick{0} &          \qswap &          \qswap &          \qswap &
          \qswap &          \qswap & \rstick{a_{4}} \qw \\
   \lstick{a_{4}} &             \qw &             \qw &             \qw &
      \qw & \qswap \qwx[-1] & \rstick{a_{3}} \qw \\
   \lstick{a_{3}} &             \qw &             \qw &             \qw &
      \qswap \qwx[-2] &             \qw & \rstick{a_{2}} \qw \\
   \lstick{a_{2}} &             \qw &             \qw & \qswap \qwx[-3] &
      \qw &             \qw & \rstick{a_{1}} \qw \\
   \lstick{a_{1}} &             \qw & \qswap \qwx[-4] &             \qw &
      \qw &             \qw & \rstick{a_{0}} \qw \\
   \lstick{a_{0}} & \qswap \qwx[-5] &             \qw &             \qw &
      \qw &             \qw &     \rstick{0} \qw \\
}$$
It should be simple to see how this can be generalised to an $n$-bit
implementation.

\subsection{Controlled Doubling Block $\braces{{\times 2}_{C}}$}
The controlled doubling block can be constructed simply by attaching all
swaps to the control.

\section{Restricted Modulo Block $\braces{\%}$}
Consider a modulo operation $a\% N$ for the special case of $0\le a<2N$. In 
this case, a single controlled subtraction is necessary in order to calculate 
the result: if $a\ge N$, subtract $N$ from $a$; otherwise, return $a$ as it 
is. This case proves sufficient for constructing the modular exponentiation 
block \cite{Markov}.

Using the ancillary qubit from a CDKM greater-equal comparator as the control 
bit in a controlled CDKM subtractor gives the desired modulo schematic:

$$\fastqc{
   \lstick{A} & \multigate{2}{\ge} & \multigate{1}{-_{C}}& \rstick{A\% B} \qw\\
   \lstick{B} &        \ghost{\ge} &        \ghost{-_{C}}&     \rstick{B} \qw\\
   \lstick{0} &        \ghost{\ge} &            \ctrl{-1}&\rstick{A\ge B} \qw\\
}$$

The modulus operation requires a single ancillary qubit, which will contain 
the value of $A\ge B$.

\section{Specific Modular Calculation Blocks}
\subsection{Modular Addition $\braces{+\%}$}
This modulo operation only works for cases when $0 \le A < 2N$; however, it 
will soon become apparent that only this case is necessary to perform modular 
exponentiation. First, consider the modular addition of two numbers $A,B
\in\integers_{N}$. By the definition of $\integers_{N}$, $0\le A,B < N$. 
Thus, $0\le A+B < 2N$; as such, the modulus part of the operation can be 
performed by the simple modulo block above, giving the modular adder as

$$\fastqc{
   \lstick{A} & \multigate{1}{+} &          \qw & \multigate{3}{\%}
      &   \rstick{A+B\%N} \qw \\
   \lstick{B} &        \ghost{+} & \push{B} \qw & \\
   \lstick{N} &              \qw &          \qw &        \ghost{\%}
      &        \rstick{N} \qw \\
   \lstick{0} &              \qw &          \qw &        \ghost{\%}
      & \rstick{A+B\ge N} \qw
}$$

However, this block leaves an ancillary qubit to contain $A+B\ge N$. As
demonstrated in \cite{beauregard2002}, it is possible to clear this bit, by
using the fact that $A+B\%N<B$ if and only if $A+B\ge N$. Thus if we invert
the ancilla when $A+B\%N<B$, it is restored to zero. Recall now that our $CMB$
block will invert a qubit if the second input is greater
than the first; as below, it does so if $B>A+B\%N$, which is what is needed to
clear the ancilla.

$$\fastqc{
   \lstick{A} & \multigate{1}{+} &          \qw & \multigate{3}{\%}
      &   \push{A+B\%N} \qw & \multigate{3}{CMB} & \rstick{A+B\%N} \qw\\
   \lstick{B} &        \ghost{+} & \push{B} \qw &
      &            \push{B} &        \ghost{CMB} & \rstick{B} \qw\\
   \lstick{N} &              \qw &          \qw &        \ghost{\%}
      &        \push{N} \qw & \\
   \lstick{0} &              \qw &          \qw &        \ghost{\%}
      & \push{A+B\ge N} \qw &        \ghost{CMB} & \rstick{0} \qw
}$$

This gives us the modular addition block
$$\fastqc{
   \lstick{A} & \multigate{3}{+\%} & \rstick{A+B\%N} \qw\\
   \lstick{B} &        \ghost{+\%} &      \rstick{B} \qw\\
   \lstick{N} &        \ghost{+\%} &      \rstick{N} \qw\\
   \lstick{0} &        \ghost{+\%} &      \rstick{0} \qw\\
}$$

\subsubsection{Controlled Modular Addition $\braces{{+\%}_{C}}$}
Simply replace the addition block with a controlled addition block. When the
control is off, the addition block is disabled; as we know that $A<N$, the
modular block will not do anything with the addition block disabled.

\subsection{Modular Doubling $\braces{\times2\%}$}
The other important operation to perform is a specific, simpler case of 
modular addition: the modular multiplication of $0\le A < N$ by 2. Similarly 
to the above, we know that $0\le 2A < 2N$, and the modulus can be taken by 
the simple modulo block. Thus, we can achieve the naive modular doubling
operation by means of the doubling block constructed earlier and the naive
implementation of the modular adder above:

$$\fastqc{
   \lstick{A} & \gate{\times2} & \multigate{2}{\%} &  \push{2A\% N} \qw\\
   \lstick{N} &            \qw &        \ghost{\%} &       \push{N} \qw\\
   \lstick{0} &            \qw &        \ghost{\%} & \push{2A\ge N} \qw
}$$

Unlike the above case, we cannot eliminate the ancillary bit through algebra.
Specifying $2A\%N$ and $N$ is not sufficient to find $A$ even if we know $A<N$;
as one counterexample, $N=4$ and $2A\%N=2$ yields $A\in\left\{1,3\right\}$,
but no specific value. Thus, it is necessary in the most general case, where
the inputs are not known classically, to have an ancillary bit containing
$2A\ge N$. In the case that $A$ and $N$ are known classically, it is a simple
matter to clear this bit; it will be shown later that we will be able to do so.

\subsubsection{Controlled Modular Doubling $\braces{{\times 2\%}_{C}}$}
Simply replace the $\braces{\times2}$ block with its controlled counterpart;
as $A<N$ the modulo gate will not do anything once the multiplier is disabled.

\subsection{Modular Multiplication $\braces{\times\%}$}
Using the $+\%$ and $\times 2\%$ blocks, we can construct a block that takes
$A,B,N$, and returns $A+B\%N,2B\%N,N$, as well as two ancillary bits which 
will contain $0$ and $2B\ge N$:
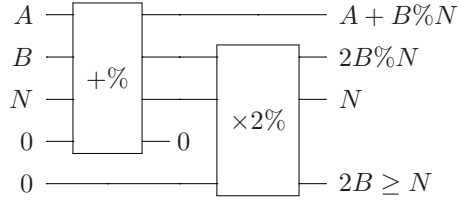
$$\fastqc{
   \lstick{A} &  \multigate{3}{+\%} & \qw & \qw & \rstick{A+B\%N} \qw\\
   \lstick{B} &         \ghost{+\%} &                 \qw &
      \multigate{3}{\times2\%} &   \rstick{2B\%N} \qw \\
   \lstick{N} &         \ghost{+\%} &                 \qw &
      \ghost{\times2\%} &       \rstick{N} \qw \\
   \lstick{0} &         \ghost{+\%} &      \push{\,\,0} \qw & &\\
   \lstick{0} &                 \qw &                 \qw &
      \ghost{\times2\%} & \rstick{2B\ge N} \qw
}$$

Making the adder in the $A+B\%N$ block controlled allows this block to 
perform a controlled addition of two numbers alongside the doubling:
$$\fastqc{
   \lstick{A} &  \multigate{3}{+\%} & \qw & \qw & \rstick{A+xB\%N} \qw \\
   \lstick{B} &         \ghost{+\%} &                  \qw &
      \multigate{3}{\times2\%} &   \rstick{2B\%N} \qw \\
   \lstick{N} &         \ghost{+\%} &                \qw &
      \ghost{\times2\%} &       \rstick{N} \qw \\
   \lstick{0} &         \ghost{+\%} &   \push{\,\,0} \qw \\
   \lstick{0} &                 \qw &                \qw &
      \ghost{\times2\%} & \rstick{2B\ge N} \qw \\
   \lstick{x} &           \ctrl{-2} &                \qw &
      \qw &       \rstick{x} \qw
}$$

Consider lining up $n$ of these blocks up, indexed 0 to $n-1$, such that the
$k$th block would have the inputs and outputs
$$\fastqc{
       \lstick{S_{k}} & \multigate{3}{+\%} & \qw & \qw
         & \rstick{S_{k+1}=S_{k}+2^{k}xA\%N}
         \qw \\
   \lstick{2^{k}A\%N} &        \ghost{+\%} &                                
      \qw & \multigate{3}{\times2\%} &   \rstick{2^{k+1}A\%N} \qw \\
           \lstick{N} &        \ghost{+\%} &                                
            \qw &        \ghost{\times2\%} &             \rstick{N} \qw \\
           \lstick{0} &        \ghost{+\%} &           \push{\,\,0} \qw \\
           \lstick{0} &                \qw &                                
            \qw &        \ghost{\times2\%} &
               \rstick{2\parens{2^{k}A\%N}\ge N} \qw \\
       \lstick{x_{k}} &          \ctrl{-2} &                                
         \qw &                      \qw &         \rstick{x_{k}} \qw
}$$
where $S_{0}=0$ and $x_{k}$ denotes the $k$th bit of an $n$-bit number $X$.
While the adders can share an ancillary bit, the doublers each require their
own bit in the general case. If $2^{k}A\forall 0\le k\le n$ and $N$ are not
known classically, this thus requires $n+1$ qubits; if these values are known,
the doublers can share their ancillary bit as well, making only 2 qubits
required.

Without the modulus operation, the resulting sum from all of these blocks is
$$\sum_{k=0}^{n-1} 2^{k}Ax_{k} = A\sum_{k=0}^{n-1} 2^{k}x_{k} = AX.$$ As
addition and multiplication are well-defined modulo $N$, this is in the same
congruence class modulo $N$ as the product with the modulus operations; thus,
the chain of blocks defines the basic modular multiplication block
$$\fastqc{
   \lstick{0} & \multigate{4}{{\times\%}_{B}} & \rstick{AX\% N} \qw \\
   \lstick{A} &        \ghost{{\times\%}_{B}} & \rstick{2^{n+1}A\% N} \qw\\
   \lstick{X} &        \ghost{{\times\%}_{B}} & \rstick{X} \qw \\
   \lstick{N} &        \ghost{{\times\%}_{B}} & \rstick{N} \qw \\
   \lstick{\mbox{$n^{*}$+1 ancilla}} &   \ghost{{\times\%}_{B}} &
                   \rstick{\mbox{$n^{*}$+1 ancilla}} \qw
}$$
where the mark $n^{*}$ is used to indicate the difference in ancilla required
if $2^{k}A$ and $N$ are or are not classically known. In the case where no
values are known classically, the multipliation requires five full registers;
in the case where all $2^{k}A$ and $N$ are known classically, it requires only
2 ancillary bits and four registers, but two of the registers are used to store
classical values. Thus, to multiply some quantum register $X$ by a known
classical value $A$ modulo a classical value $N$ requires $2n+4$ qubit
registers, all of which contain results after the operation finishes.

Note that with the basic block, an empty register is required at the outset,
but no empty register is provided at the close; this could pose an issue when
laying blocks in succession. However, when $N$ is odd and $A$ is coprime to
$N$ (both of can be assumed for Shor's algorithm), a solution to this issue
presents itself. When $A$ and $N$ are both classically known, it is a simple
matter to determine $A^{-1}$ modulo $N$ classically prior to the algorithm;
when $A$ is not classically known, \cite{proos2004} details an
$O\parens{n^{2}}$ implementation of the quantum extended Euclidean algorithm
for coprime $A$ and $N$, which can likewise determine $A^{-1}$.

We know that $N$ is odd (if $N$ were even it would have a trivial factor of
2, making Shor's algorithm unnecessary), and thus $2^{n+1}$ has an inverse
modulo $N$; as both $N$ and $2^{n+1}$ are classically knowable, we can
preconstruct a circuit to multiply by $\parens{2^{n+1}}^{-1}$ mod $N$;
no information is lost, as multiplying by $2^{n+1}$ reverses it.

We now construct the circuit
$$\fastqc{
   \lstick{0} & \multigate{3}{{\times\%}_{B}}
      & \qw
      & \qw
      & \push{AX\% N} \qw
      & \qswap \qwx[2]
      & \multigate{3}{{\times\%}_{B}}
      & \rstick{0} \qw\\
   \lstick{A} &        \ghost{{\times\%}_{B}}
      & \gate{\times \parens{2^{n+1}}^{-1}\% N}
      & \multigate{2}{{}^{-1}\%}
      & \push{A^{-1}\% N} \qw
      & \qw
      & \ghost{{\times\%}_{B}}
      & \rstick{2^{n+1}A^{-1}\% N} \qw\\
   \lstick{X} &        \ghost{{\times\%}_{B}}
      & \rstick{X} \qw
      & 
      & \lstick{X}
      & \qswap
      & \ghost{{\times\%}_{B}}
      & \rstick{AX\% N} \qw\\
   \lstick{N} &        \ghost{{\times\%}_{B}}
      & \qw 
      & \ghost{{}^{-1}\%}
      & \push{N} \qw
      & \qw
      & \ghost{{\times\%}_{B}}
      & \rstick{N} \qw
}$$
After the last multiplication, the top register holds $X\oplus A^{-1}AX\% N$,
which clearly cancels to 0 in all cases. It is a simple matter of again
multiplying by $\parens{2^{n+1}}^{-1}$ then reversing the modular inverse gate
(which has the added benefit of clearing its ancilla) to yield the modular
multiplication block
$$\fastqc{
   \lstick{0} & \multigate{3}{\times\%} & \rstick{0} \qw\\
   \lstick{A} &        \ghost{\times\%} & \rstick{A} \qw\\
   \lstick{X} &        \ghost{\times\%} & \rstick{AX\%N} \qw\\
   \lstick{N} &        \ghost{\times\%} & \rstick{N} \qw\\
}$$
This block contains two basic modular multiplications which each use $n^{*}+1$
ancillary bits, though the $+1$ can be recycled, leaving $2n^{*}$ uncleared
ancilla. There are two modular multiplication blocks and two modular inverse
blocks, all with performance $O\parens{n^{2}}$, so this block also is
$O\parens{n^{2}}$

\subsubsection{Controlled Modular Multiplication $\braces{{\times\%}_{C}}$}
There are multiple ways to construct a controlled multiplication block. The
most obvious means is to simply replace all of the adders with a controlled
version. Alternately, adding an additional register and two sets of $n+1$
Fredkin gates gives
$$\fastqc{
   \lstick{1} & \qswap \qwx[2] & \qw & \qswap \qwx[2] & \rstick{1} \qw\\
   \lstick{0} &    \qw & \multigate{4}{\times\%} & \qw & \rstick{0} \qw\\
   \lstick{A} & \qswap & \ghost{\times\%} & \qswap
      & \rstick{A}\qw\\
   \lstick{X}    & \qw & \ghost{\times\%} & \qw & \rstick{A^{c}X\% N} \qw \\
   \lstick{N}    & \qw & \ghost{\times\%} & \qw & \rstick{N} \qw \\
   \lstick{\mbox{$2n^{*}$ ancilla}}
      & \qw & \ghost{\times\%} & \qw & \rstick{\mbox{$2n^{*}$ ancilla}} \qw\\
   \lstick{c} & \ctrl{-4} & \qw & \ctrl{-4} & \rstick{c} \qw
}$$
This construction reduces the number of controls needed, which reduces the
complexity of the overall circuit. It even does not impact the number of
ancillary bits; simply make the clearing of the bit after each doubling be
controlled by the control bit as well, and clear based on classically-known
values if the control states that the multiplication is being performed.

There is no clear advantage of one control implementation over another; the
naive implementation requires more complex gates, and introduces a large
number of controls, while the Fredkin implementation requires an additional
full register of $n+1$ qubits. In the modular exponentiation case, we will
omit the register needed for the Fredkin implementation.

\subsubsection{Alternate Modular Multiplication Blocks}
Markov and Saeedi\cite{Markov} construct several multiplication blocks for
specific values of $A$ and/or $N$, which offer significant advantages to the
circuit above in exchange for loss of generality. They also propose a distinct
implementation of the modular multiplication block which uses a
division-with-remainder method, as opposed to the repeated-addition method
proposed above. This requires $\floor{\log_{2} A}+n+1$ ancillary bits, more
than the circuit proposed above, but offers significant advantages for some
values of $A$ and $N$, particularly when $A^{2}<N$.

\section{Modular Exponentiation for Shor's Algorithm}
In our implementation of Shor's algorithm, we wish to take a non-classically
determined value $A$ and raise it to a superposition of exponents modulo $N$,
where $N$ is an input parameter (and thus classically known). We will do so by
repeated multiplication by $A^{2^{k}}\%N$ for $0\le k\le n$, using $n+1$
modular multiplication blocks.

The $k$th multiplication block is of the form
$$\fastqc{
   \lstick{0}
      & \multigate{3}{{\times\%}_{C}}
      & \rstick{0} \qw\\
   \lstick{A^{2^{k}}}
      & \ghost{{\times\%}_{C}}
      & \rstick{A^{2^{k}}} \qw\\
   \lstick{P_{k}}
      & \ghost{{\times\%}_{C}}
      & \rstick{P_{k+1} = P_{k}A^{2^{k}y_{k}}\% N} \qw\\
   \lstick{N}
      & \ghost{{\times\%}_{C}}
      & \rstick{N} \qw\\
   \lstick{y_{k}}
      & \ctrl{-1}
      & \rstick{y_{k}} \qw
}$$
where $P_{0}=1$. It is then necessary to square $A^{2^{k}}$ modulo $N$ between
multiplication blocks, as we are working with $A$ not classically known.

For this we will use a naive implementation requires an additional
$n+1$-qubit register into which $A^{2^{k}}\%N$ is copied using CNOT gates,
followed by an additional modular multiplication, which thus requiresa totoal
of $3n+2$ qubits, one of which is recyclable prior to the end of the
calculation. More effective quantum circuits for modular square are not
considered in this paper, and none were readily found in literature;
implementations of Shor's algorithm for general $N$ focus on classically known
$A$, as $A$ can be easily determined classically in the typical case. Thus,
our consideration of this is largely to demonstrate the advantages of a
partially-classical circuit over a purely quantum one.

Modular exponentiation of classically unknown $A$ thus requires five input
registers and $3n^{2}+n+1$ ancillary qubits, for a total of $3n^{2}+6n+6$
qubits. In the case where $A$ is classically known, the modular multiplication
blocks do not have any unclearable ancillary bits, requiring a total of 1
ancillary bit each, and it is not necessary to perform a quantum modular
square at all ($A^{2^{k}}\%N$ can be found classically for all $k$), reducing
the number of qubits needed to $5n+6$ (one ancillary bit and five registers).

\section{Conclusion}
We construct an $n$-bit implementation of modular exponentiation $A^{x}\%N$
which only requires that $N$ be classically known, and the corresponding
implementation where $A$ and $N$ are both classically known. This circuit uses
$O\parens{n^{3}}$ gates, or equivalently $O\parens{\parens{\log N}^{3}}$
gates; this concurs with theoretical expectations of the period-finding routine.
When $A$ is classically known, the number of ancillary qubits drops off and
reduces to be comparable to other general-$N$ implementations of Shor's
algorithm. This requires a general case of $3n^{2}+6n+6$ qubits, which can be
reduced to $5n+6$ in the event of a classically-known value of $A$ (assuming
that modifications can be made to the circuit based on that classical value),
clearly illustrating the advantages of classical control over $A$. This
$O\parens{\parens{\log N}^2}$ gate count is the best we have found for
classically unknown $A$, while $O\parens{n}$ is the minimum necessary to
perform Shor's Algorithm on general $N$. The circuit also operates in
$O\parens{n^{3}}$ depth, as it is not parallellised.

There are numerous variations on the circuit, which offer various advantages
and disadvantages. Table \ref{tab:comparison} lists a comparison of our
construction with other implementations of modular exponentiation; in most
cases, it offers situational advantages over other implementations: it has more
effective asymptotic behaviour in one area and worse in another. While
\cite{beckman1996} and \cite{vedral1996} offer similar asymptotic behaviour to
our own circuit, the former is designed for ion-trap hardware and the latter
requires more non-asymptotic qubits.

\begin{table}
   \begin{tabular}{|c|c|c|c|}
   \hline
      Implementation
         & Depth
         & Gates
         & Qubits\\ \hline
      \cite{beauregard2002}
         & $O\parens{n^{3}}$
         & $O\parens{n^{3}\log n}$
         & $O\parens{n}$\\ \hline
      \cite{pavlidis2013}
         & $O\parens{n^{2}}$
         & $O\parens{n^{3}}$
         & $O\parens{n}$\\ \hline
      \cite{pham2013}
         & $O\parens{\log^{2} n}$
         & $O\parens{n^{4}}$
         & $O\parens{n^{4}}$\\ \hline
      \cite{meter2005}
         & $O\parens{n^{2}\log n}$
         & $O\parens{n^{3}}$
         & $O\parens{n^{2}}$\\ \hline
      \cite{kutin2006}
         & $O\parens{n^{2}}$
         & $O\parens{n^{3}}$
         & $O\parens{n}$\\ \hline
      \cite{zalka2008}
         & $O\parens{n^{2}}$
         & $O\parens{n^{3}}$
         & $O\parens{n}$\\
      (two algorithms)
         & $O\parens{n}$
         & $O\parens{n^{2}}$
         & $O\parens{n^{1.2}}$\\ \hline
      \cite{fowler2004}
         & $O\parens{n^{3}}$
         & $O\parens{n^{4}}$
         & $O\parens{n}$\\ \hline
      \cite{beckman1996}
         & $O\parens{n^{3}}$
         & $O\parens{n^{3}}$
         & $O\parens{n}$\\ \hline
      \cite{vedral1996}
         & $O\parens{n^{3}}$
         & $O\parens{n^{3}}$
         & $O\parens{n}$\\ \hline
      Current Work
         & $O\parens{n^{3}}$
         & $O\parens{n^{3}}$
         & $O\parens{n}$\\ \hline
   \end{tabular}
   \caption{Asymptotic comparison of select Shor's Algorithm implementations}
   \label{tab:comparison}
\end{table}

The number of gates required is linear for a single adder and the derived
blocks, quadratic for multiplication, and cubic for exponentiation, which is
in keeping with the general circuits discussed in passing by \cite{Markov} in
comparison to their linear specific-case circuits.

There are two clear avenues for further work: first, developing a means to
square a number modulo $N$ (a non-reversible operation by itself) on fewer
than $n+1$ ancillary qubits and in less than $O\parens{n^{2}}$ time should be
possible, but no implementation was discovered as this paper was being
written; all reviewed implementations of Shor's algorithm using repeated
modular multiplication focused on classically-known values of $A$. This is
largely a matter of curiosity, to determine how a quantum $A$ might work, as
$A$ can be determined classically rather simply.

Further, we do not consider the impact of error correction on our circuit,
instead focusing on simply constructing the framework. An actual implementation
would likely require error correction, and thus it would be necessary to
determine any consequences that an appropriate scheme might have on the design
and performance of the circuit.

While compiled circuits can use much smaller numbers of qubits, and are thus
often more useful for experimental tests, they lose out on generality. A
general formulation gives a broader view of the problem which the compiled
circuits may overlook, and is necessary for any practical realisation of the
algorithm in the future. Our proposed general implementation of Shor's
algorithm provides a blueprint for large-scale quantum photonic integrated
circuit realizations. 

\bibliography{shor}

\end{document}